\documentclass[prl,letterpaper,twocolumn,showpacs,superscriptaddress,floatfix]{revtex4}

\usepackage{graphicx,psfrag,amsmath,amssymb,amsfonts,bbm,latexsym,color,dcolumn,bm}

\begin{document}

\title{Anomalies in electrostatic calibrations for the \\
  measurement of the Casimir force in a sphere-plane geometry}

\author{W.J. Kim$^*$}
\affiliation{Department of Physics and Astronomy,Dartmouth
  College,6127 Wilder Laboratory,Hanover,NH 03755,USA}

\author{M. Brown-Hayes}
\affiliation{Department of Physics and Astronomy,Dartmouth
  College,6127 Wilder Laboratory,Hanover,NH 03755,USA}

\author{D.A.R. Dalvit}
\affiliation{Theoretical Division, MS B213, Los Alamos National Laboratory,
Los Alamos, NM 87545, USA}

\author{J.H. Brownell}
\affiliation{Department of Physics and Astronomy,Dartmouth
  College,6127 Wilder Laboratory,Hanover,NH 03755,USA}

\author{R. Onofrio}
\affiliation{Dipartimento di Fisica ``Galileo Galilei'', Universit\`a  di Padova,
Via Marzolo 8, Padova 35131, Italy}

\affiliation{Department of Physics and Astronomy,Dartmouth
  College,6127 Wilder Laboratory,Hanover,NH 03755,USA}

\date{\today}

\begin{abstract}
We have performed precision electrostatic calibrations in the
sphere-plane geometry, and observed anomalous behavior.
Namely, the scaling exponent of the electrostatic signal with distance was
found to be smaller than expected  on the basis of the pure Coulombian
contribution, and the residual potential found to be distance
dependent. We argue that these findings affect the accuracy of the
electrostatic calibrations and invite reanalysis of previous
determinations of the Casimir force.

\end{abstract}


\maketitle
Over the last decades the Casimir force \cite{Casimir} has met
increasing popularity as a macroscopic manifestation of quantum
vacuum \cite{Milonni}, with its relevance spanning from nanotechnology to
cosmology \cite{Bordag}. With the claimed accuracy of recent
experiments ranging from 15$\%$ in the parallel plane case
\cite{Bressi} to 0.1-5$\%$ in the sphere-plane case
\cite{Lamoreaux,Mohideen,Chan,Decca}, it also provides constraints 
on the existence of forces superimposed to the Newtonian gravitational 
force and expected in various unification attempts \cite{Fishbach,Onofrio}.

Concern has been raised that previous analyses in the measurement 
of the Casimir force have overlooked possible influence of residual 
electric effects (the so-called patch effects), which could mimic 
the Casimir force \cite{Speake}.
Residual electric effects are known to play an important role in the
measurement of van der Waals force between macroscopic bodies,
where corrections based on a model for work function anisotropies and
their associated patch charges have been discussed \cite{Anisotropy}.
Some of the early studies focusing on various adhesion and friction
surface forces can be found in \cite{Trappedcharge,Electrification,Blackman}.
More recently, Stipe {\it et al.} \cite{Stipe} argue that the presence
of an inhomogeneous tip-sample electric field in an AFM type
experiment is difficult to avoid, thereby imposing a significant
limitation to the accuracy of force measurements.
This issue indicates a common problem in electrostatic measurements 
over a range of different experimental circumstances and distance 
scalings, and calls for more attention on calibration procedures, 
as the accuracy claimed in Casimir force measurements inherently 
depends on the quality of the corresponding electrostatic calibrations. 

We report here an electrostatic calibration procedure for a
sphere-plane geometry in which the electric signal is studied
at all explored distances. Our results are obtained in a range of parameters
that interpolates between the two previous sets of sphere-plane measurements.
We use a spherical lens with a large radius of curvature, similar to the
experiment performed in \cite{Lamoreaux}, while at the same
time exploring distances down to few tens of nanometers from the
point of contact between the sphere and the plane, similar to
more microscopic setups using microresonators \cite{Mohideen,Chan,Decca}.
The measurements reveal anomalous behavior not reported to date.

\begin{figure}[t]
\begin{center}
\includegraphics[width=0.70\columnwidth]{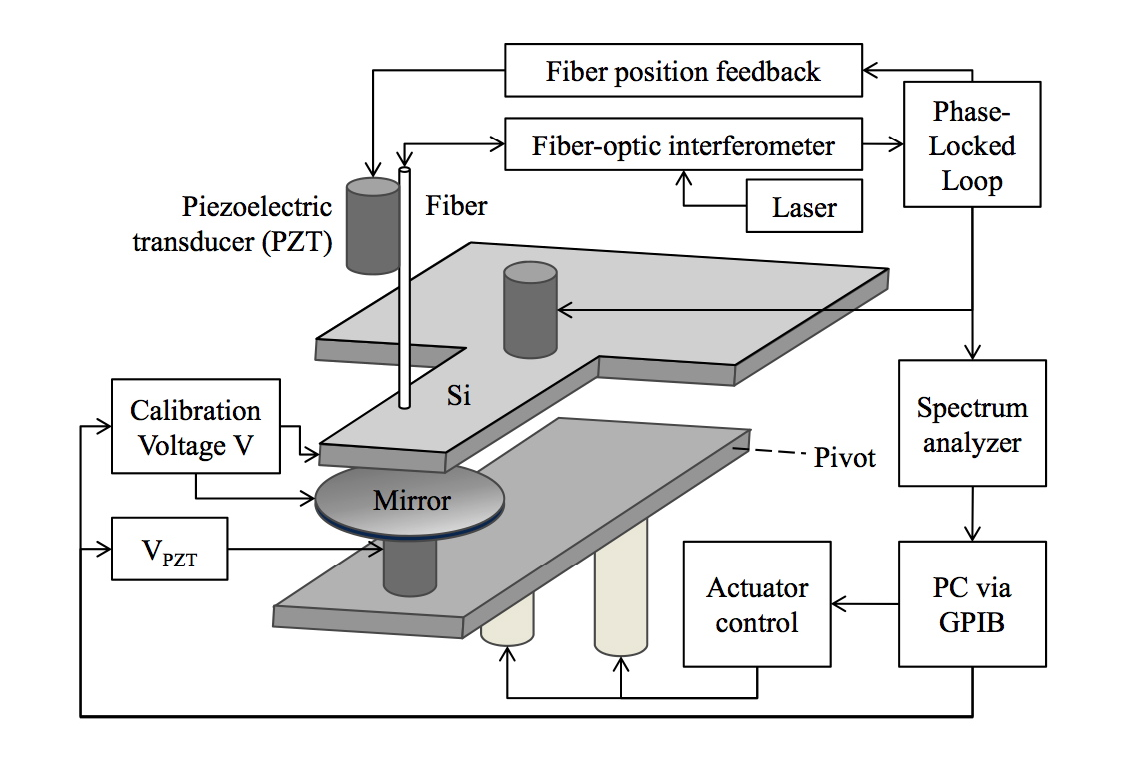}
\end{center}
\caption{(Color online) Schematic of the experimental setup. The silicon resonator,
  of length $L=(22.56 \pm 0.01)$ mm, width $w=(9.93\pm 0.01)$ mm, and
  thickness $t=(330 \pm 10)$ $\mu$m with the physical mass $m_p=(1.72
  \pm 0.05) \times 10^{-4}$ kg, is opposed by a spherical mirror with
  radius of curvature $R=(30.9 \pm 0.15)$ mm and diameter $a=(8.00 \pm
  0.25)$ mm. The mirror is mounted on an aluminum frame connected to
  two motorized actuators allowing for coarse translational motion,
  plus an additional piezoelectric transducer for fine translational
  motion driven with a bias $V_{\mathrm{PZT}}$. Both the spherical
  mirror and the surface of the cantilever are coated by thermal evaporation
  with a 200 nm layer of gold with 1-2 nm rms roughness.}
\label{prarc.fig1}
\end{figure}

Our experimental setup is an upgrade of the previous arrangement for
measuring the Casimir force in the cylinder-plane configuration
\cite{Brown}. Force gradients between a spherical lens and a silicon
cantilever are detected by measuring
the shift in the mechanical oscillation frequency of the cantilever.
A schematic of the apparatus is shown in Fig. \ref{prarc.fig1}.
The cantilever is electrically isolated and thermally stabilized by a
Peltier cooler to within 50 mK. The motion of the cantilever is
monitored by a fiber optic interferometer \cite{Rugar1} positioned a
few tens of microns above the cantilever. The output signal from the
interferometer is put through a single reference mode lock-in
amplifier and is fed back into the piezoelectric actuator driving the
cantilever motion, forming a phase-locked loop (PLL) with an optimized
phase angle around 30-40 degrees \cite{Albrecht}.
The measured frequency at a typical vacuum pressure of
$1.6 \times 10^{-4}$ Torr is consistent with the
predicted frequency of the fundamental flexural mode of the
cantilever, around 894 Hz. The stiffness $k$ of the resonator is
estimated to be $5.4 \times 10^3$ N/m, $10^4$ to $10^5$ times higher
than that of typical cantilevers used in atomic force microscopy.
Our cantilever flexes at most 1 nm, allowing very  small gaps
to be probed, at the cost of an overall lower force sensitivity
that limits the maximum explorable distance.
No compensating external voltages are needed to prevent snapping.

In general, the square of the measured frequency $\nu_{m}$ of a
cantilever under the influence of generic forces is
\begin{equation}
\label{generic}
\nu_{m}^2(d, V)=\nu_p^2(d\rightarrow\infty)-\Delta\nu_e^2(d, V)-\Delta\nu_r^2 (d),
\end{equation}
where $\nu_p$ is the cantilever's natural flexural frequency,
$\Delta\nu_e^2$ is the frequency shift due to externally
applied voltages $V$ at different gap separations $d$,
and $\Delta\nu_r^2$ is the frequency shift subject to distance-dependent
forces of non-electrostatic nature, for instance the Casimir
force.  For the sphere-plane configuration and for our choice of
parameters, the proximity force approximation (PFA)
\cite{Blocki} holds, the electric force gradient is
$F'_{\rm el}=\pi\epsilon_0RV^2/d^2$, and Eq. (\ref{generic}) can be
parameterized in the presence of a contact potential $V_c$ as
$\nu_{m}^2=\nu_0^2-k_{\mathrm{el}}(V-V_c)^2$,
where $\nu_0^2=\nu_p^2-\Delta\nu_r^2$ , a parabola whose maximum is
reached when the applied voltage equals to $V_c$. The parabola
curvature $k_{\mathrm{el}}=\epsilon_0R/4\pi m_{\mathrm{eff}}d^2$
reflects the cantilever response to externally applied electric forces
at a given distance. All three parameters, $\nu_0^2, k_{\mathrm{el}},$
and $V_c$ are simultaneously evaluated for every step in a data sequence
and so they can be plotted as a function of gap separation.
In what follows, we scrutinize these parameters and discuss the systematic
issues that must be addressed to validate a consistent analysis of the
Casimir force.
The value of $k_{\mathrm{el}}$ as a function of distance directly
characterizes the system's response to an applied bias and sets the
basis for  residual distance-dependent force analysis. 
The gap distance cannot be known with sufficient accurately
prior to a force measurement \cite{Onofrio}. In a typical experiment,
the gap is varied by the voltage applied to the PZT
($V_{\mathrm{PZT}}$) and consequently $k_{\mathrm{el}}$ is a function
of relative distance (i.e. of the applied $V_{\mathrm{PZT}}$).
This requires an additional fitting parameter $V_{\mathrm{PZT}}^0$, which would
cause contact and must be inferred from fitting the function
\begin{equation}
\label{curvature}
k_{\mathrm{el}}(V_{\mathrm{PZT}})={\alpha}{(V^0_{\mathrm{PZT}}-V_{\mathrm{PZT}})^{-2}},
\end{equation}
where $\alpha$ is a calibration factor containing the effective mass
of the cantilever through $\alpha\equiv\epsilon_0R/4\pi
m_{\mathrm{eff}} \beta^2$, and $\beta$ is the conversion factor translating
the PZT voltage to the actual distance unit.
\begin{figure}[t]
\includegraphics[width=0.80\columnwidth]{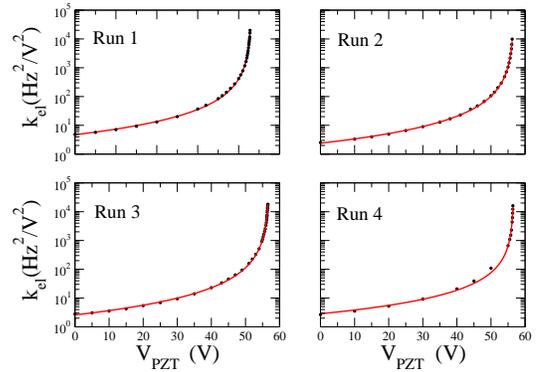}
\caption{(Color online) Curvature coefficients data $k_{\mathrm{el}}$
versus applied voltage $V_{\mathrm{PZT}}$ and best fit with Eq.
(\ref{curvature}), but leaving the exponent as a free parameter
for four different experimental runs. Given 4\% uncertainty in $k_{el}$ , the reduced 
$\chi^2$ are 1.0, 0.8, 1.2, and 7.0 with the exponent left as a free parameter for 
Run 1, Run 2, Run 3, Run 4, respectively.  Fixing the exponent to be -2 increases 
these values to be 16, 7.7, 6.9, and 37. The data follows a pure power law in 
all explored distances. The absolute distances can be assessed from the
asymptotes of the curves $V^0_{\mathrm{PZT}}$, which are, in the
progressive order of run,  $43.12\pm0.01$ V, $56.77\pm0.02$ V,
$56.99\pm0.02$ V, and $56.66\pm0.02$ V. The different values of
$V^0_{\mathrm{PZT}}$ in the four runs, taken in different weeks, are
due to the rearrangement of the sphere location obtained using
the actuator drivers.}
\label{prarc.fig2}
\end{figure}
This is common practice in most of the Casimir
force measurements in which the sensitivity of the apparatus and
the zero distance are extracted from an electrostatic calibration.
Depending on apparatus type, extracted quantities can be
either spring constant \cite{Mohideen}, torsion constant
\cite{Lamoreaux,Chan,Decca}, or effective mass like in this case
\cite{Bressi,Brown}. Note that the absolute distance can be expressed
in two ways: $d(V_\mathrm{PZT})=\beta(V_{\mathrm{PZT}}^0-V_{\mathrm{PZT}})$ or
$d(V_\mathrm{PZT})=\beta{[\alpha/k_{\mathrm{el}}(V_\mathrm{PZT})]}^{1/2}$. Therefore, the absolute
distance can be inferred either from the
asymptotic limit $V^0_{\mathrm{PZT}}$ of the fit function or from the
calibration factor $\alpha$ of the same function, indicating an
interdependency of  the two physical parameters appearing in
Eq. (\ref{curvature}).

The above procedure and all subsequent analysis is inapplicable if the
data fail to follow the inverse square law of Eq. (\ref{curvature}).
Surprisingly, our experimental data from four separate sequences
follow a power law similar to Eq. (\ref{curvature}), but with
exponents  $-1.70\pm0.01, -1.77\pm0.02, -1.80\pm0.01,
-1.54\pm0.02$, far from the expected value of -2.
Figure \ref{prarc.fig2} shows plots of $k_{\mathrm{el}}$
as a function of  $V_{\mathrm{PZT}}$, with the exponent
left as a free parameter.
\begin{figure}[t]
\includegraphics[width=0.75\columnwidth]{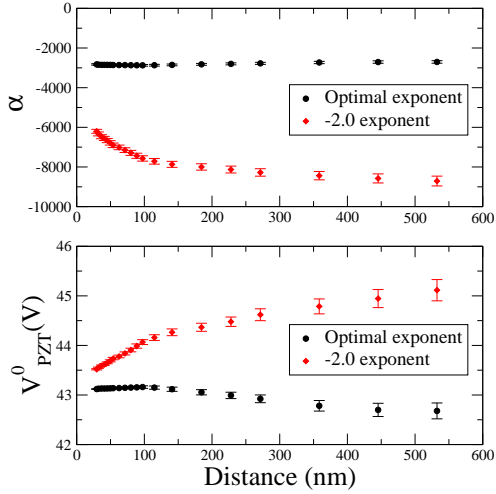}
\caption{(Color online) Stability test of  fit parameters in Eq.~(\ref{curvature})
for Run 1 with the exponent taken to be -1.7 (black, circles) and -2
(red, diamonds). Both calibration factor $\alpha$ (top plot) with units
of Hz$^2$ V$^{-2}$ m$^{1.7}$ for the -1.7 exponent, and Hz$^2$ V$^{-2}$ m$^{2}$
for the -2.0 exponent, and the zero distance bias $V^0_{\mathrm{PZT}}$
(bottom plot) remain stable within 3\% in the flat region at the closest
separation for the exponent -1.7, while a large spreading in the
two parameters is evident for the exponent -2. Note also the
correlation between the fit parameters reflecting their
interdependency as discussed in the text.}
\label{prarc.fig3}
\end{figure}
With the apparatus stationary, we observe a random measurement uncertainty of 
4\% in $k_{el}$ superimposed on a long time scale (comparable to the duration of 
one data run) drift, which may be due to a systematic, thermally induced change 
in gap size up to $\pm 200$ nm.  Given the 4\% error, the reduced $\chi^2$ is 
near one when the exponent is fit in the -1.7 to -1.8 range compared to around 
10 for fixed -2.0 exponent.  Adherence of our data to a strict power law from 
the farthest to the closest approach in all four cases implies the drift was 
not severe during our data runs, except possibly in the case of Run 4. 
To check the stability of the electrostatic result
ruled by the unexpected power law, the following test has
been conducted. We repeat fitting the data, $k_{\mathrm{el}}$
versus $V_{\mathrm{PZT}}$, starting with few points at the largest
distances and by progressively including the data point
corresponding to the closer distances. 
In this way the intrinsic instability of the fit will 
manifest itself through a systematic variation
in the fit parameters subject to a choice of data points included.
Fig. \ref{prarc.fig3} shows the result of the stability test for Run 1.
\begin{figure}[t]
\centering
\includegraphics[width=0.90\columnwidth]{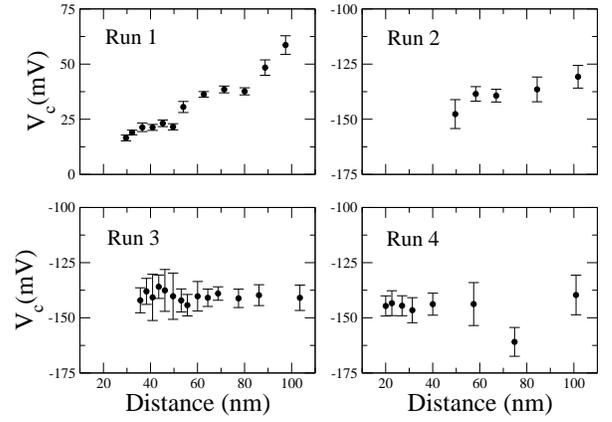}
\caption{Dependence of the contact potential versus sphere-plane
separation. Two noticeable behaviors are a linear dependence on
the separation (Run 1 and Run 2) and a settlement to a constant
contact potential at the smallest explored distances (Run 3 and Run
4). In Run 1 the uppermost point of the sphere is different from the
other, successive three runs as the latter were taken after having
tilted the sphere by 0.05 radians, to check the sensitivity to local
surface details. Note that in Run 1, $V_c$ is positive while it is negative and
converges to $V_c \simeq$ -150 mV at the smallest separations for other runs.}
\label{prarc.fig4}
\end{figure}
The electrostatic fit with the exponent of -1.7 displays a
stable region at the shortest separation gap corresponding to the
inclusion of most data points. Excluding data points simply makes the
error bar larger, while the fit parameters remain stable. Forcing the
exponent to be -2, on the other hand, leads to unsatisfactory
calibration results, as the parameters start to drift immediately from
the outset and do not display any stable region.
More than 40\% variations in both the
calibration factor and the zero distance are found in this case.
The instability is already present at the largest distances,
indicating that the source of deviation from pure Coulombian behavior,
for instance due to patch potentials, is still effective.
Runs 2, 3 and 4 exhibit similar behavior.
In order to further test the consistency of the electrostatic
calibration, we have compared the absolute distances obtained
from the two methods outlined above, getting a mutual agreement
within 2 $\%$ for the fit with the -1.7 exponent at all distances,
and deviations of 20 $\%$ if the fitting exponent is instead fixed to be -2.

The unexpected power law poses a significant limit on the validity of
our electrostatic calibration. Here, we briefly examine some
hypotheses which could potentially explain a deviation from the expected
power law. \textit{Static deflection of cantilever}: The spring
constant of our cantilever is extremely stiff (about 5400 N/m). Using
Hooke's law, a deflection experienced by the cantilever due to an
electrostatic force at 100 nm with an applied voltage of 100 mV is
less than 0.2 $\AA$. Hence, the static deflection should play little
role. \textit{Thermal drift}: Even though the temperature of the
cantilever is actively stabilized  by a Peltier cooler, the rest of
the system is still subject to global thermal variation.  In order to
see this, we have measured $k_{\mathrm{el}}$ with respect to time at a
nominally fixed distance. In the worst circumstance, the gap separation during
the course of measurements can drift as much as 200 nm in
\textit{either} direction. Although such a drift could in principle
affect the inferred exponent, a highly unlikely non-linear
monotonic drift would be necessary to account for the consistently 
observed anomaly in each independent run. \textit{Non-linearity of the PZT
 translation}: The linearity of the PZT translation has been tested
under a number of different circumstances. Notice that the translation
intervals between the data points in each of the runs shown in
Fig. \ref{prarc.fig2} are completely random. Yet, all of the runs
obey a specific power law in all distances. The PZT was also independently
calibrated by fiber optic interferometer with a consistent, linear 
actuation coefficient factor $\beta=87\pm2$ nm/V.
\textit{Non-linear oscillation of cantilever}: The cantilever is
driven at resonance in a phase-locked loop, a routine technique
adopted by many groups \cite{Decca,Albrecht,Jourdan,Giessbl}.
Higher order terms in the force expansion should produce higher
harmonics of the drive frequency. Then the assumption that the frequency
shift is simply proportional to the gradient of the external force
$F'(d)$ could lead to erroneous assessments of  $k_{\mathrm{el}}$,
eventually affecting the exponent. We have not observed higher harmonics
in the frequency spectrum of the resonator.
\textit{Surface roughness}: The deviation from geometrical ideality
and its influence on the local capacitances \cite{Boyer} could in
principle play a role especially at the smallest distances.
However with the measured rms values of roughness for the two
surfaces we find the corrections negligible, as we discuss in
detail elsewhere \cite{Kimthesis}.

After discarding the hypotheses above, we may consider the effect of
the patch surface potentials. Patch effects are expected to induce
deviations from the Coulombian scaling with distance \cite{Speake},
and we have found another anomaly corroborating this hypothesis.
Indeed, our technique for obtaining the parabola curvatures at all
distances reveals that the contact potential depends on distance, as
shown in Fig. \ref{prarc.fig4}.
In runs 1 and 2, the contact potentials appear to be distance-dependent.
This finding calls for a more careful analysis of previous experiments in which the
determinations of the contact potential have been performed at
relatively large distances, typically above 1 $\mu$m. To see this more
clearly, we evaluate the equivalent voltage
$V_{\mathrm{eq}}=(\pi/d)\sqrt{ \hbar c/360\epsilon_0}$ necessary to
mimic the Casimir force at a given distance, as first discussed in
\cite{OnoCaru}. At 1 $\mu$m, the magnitude of
the Casimir force is equivalent to that from an uncompensated voltage of 10 mV
between the two surfaces, which should be compared to variations of 90
mV and 50 mV of the measured contact potentials in Run 1 and Run 2,
respectively. The determination of the contact potential at
large distances, as usually performed in various experiments, may
therefore lead to a spurious signal of electrostatic origin
if the contact potential at smaller distances is only partly
compensated by an external counterbias.

Once the anomalous scaling exponent and the distance dependent
residual voltage are taken into account, one can look for the
distance dependence of $\nu_0^2$.
A distant-dependent frequency should
signal forces such as the Casimir force and/or forces due to
surface potentials related with patch effects \cite{Speake}.
Disentangling forces of different origins becomes exceedingly 
complex with the unexpected power law found in the electrostatic 
analysis. Preliminary residual fitting indicate that the 
exponents obtained for Run 1 and Run 3 are $-2.12 \pm 0.16$ 
and $-3.64 \pm 0.25$, respectively, systematically smaller 
than -4 expected for the Casimir force \cite{Kimthesis}.

Finally, we emphasize that  the study of the Casimir force should be
regarded as an extension of previous van der Waals force measurements
with AFM techniques, since the underlying physics governing the
short range forces between closely spaced bodies should be the same.
Some of the systematic effects discussed here have been extensively
discussed in the AFM literature \cite{Giessbl,Cappella,Garcia}.
Although we cannot draw substantial conclusions about the observation
of the Casimir force itself from our data, the experimental procedure
outlined in this report should provide an optimal strategy to handle
contact potentials at all distances. Apart from looking for confirmation
of the observed anomalies in other experimental setups, we believe that
our findings call for a reanalysis of previous Casimir force experiments
in the sphere-plane geometry, a check of the claimed accuracy, and
for the systematic control of the effect of uncompensated patch potentials.

\vspace{0.1cm}
\noindent
$^*${Present address: Department of Physics, Yale University,
  217 Prospect Street, New Haven, CT 06520-8120.}

\vspace{0.1cm}
\noindent
We are grateful to G. Carugno, H.B. Chan, S.K. Lamoreaux, J.N. Munday, and
A. Parsegian for useful discussions. We also thank R. Johnson for technical
support.


\begin{references}

\bibitem{Casimir} H.B.G. Casimir, Proc. K. Ned. Akad. Wet. B \textbf{51}, 793 (1948).

\bibitem{Milonni} P.W. Milonni, {\it The Quantum Vacuum} (Academic Press, San Diego, 1994).

\bibitem{Bordag} M. Bordag, U. Mohideen, and V. M. Mostepanenko, Phys. Rep. \textbf{353}, 1 (2001).

\bibitem{Bressi} G. Bressi {\it et al}., Phys. Rev. Lett. \textbf{88}, 041804 (2002).

\bibitem{Lamoreaux} S.K. Lamoreaux, Phys. Rev. Lett. \textbf{78}, 5 (1997).

\bibitem{Mohideen} U. Mohideen and A. Roy, Phys. Rev. Lett. \textbf{81}, 4549 (1998).

\bibitem{Chan} H.B. Chan {\it et al}., Science \textbf{291}, 1941 (2001).

\bibitem{Decca} R.S. Decca {\it et al}., Phys. Rev. Lett. \textbf{91}, 050402 (2003).

\bibitem{Fishbach} E. Fischbach and C. L. Talmadge, {\it The Search for Non-Newtonian Gravity} (AIP/Springer-Verlag, New York, 1999).

\bibitem{Onofrio} R. Onofrio, New J. Phys. \textbf{8}, 237 (2006).

\bibitem{Speake} C. C. Speake and C. Trenkel, Phys. Rev. Lett. \textbf{90}, 160403 (2003).

\bibitem{Anisotropy} N. A. Burnham, R. J. Colton, and H. M. Pollock, Phys. Rev. Lett. \textbf{69}, 144 (1992).

\bibitem{Trappedcharge} J. M. R. Weaver and D. W. Abraham, J. Vac. Sci. Techn. B \textbf{9}, 1559 (1991).

\bibitem{Electrification} B. D. Terris {\it et al}., Phys. Rev. Lett. \textbf{63}, 2669 (1989).

\bibitem{Blackman} G. S. Blackman, C. M. Mate, and M. R. Philpott, Phys. Rev. Lett. \textbf{65}, 2270 (1990).

\bibitem{Stipe} B. C. Stipe {\it et al}., Phys. Rev. Lett. \textbf{87}, 096801 (2001).

\bibitem{Brown} M. Brown-Hayes {\it et al}., Phys. Rev. A \textbf{72}, 052102 (2005).

\bibitem{Rugar1} D. Rugar, H. J. Marmin, and P. Guethner, Appl. Phys. Lett. \textbf{55}, 2588 (1989).

\bibitem{Albrecht} T. R. Albrecht {\it et al}., J. Appl. Phys. \textbf{69}, 668 (1991).

\bibitem{Blocki} J. Blocki {\it et al.}, Ann. Phys. \textbf{105}, 427 (1977).

\bibitem{Jourdan} G. Jourdan {\it et al}., E-Print arXiv:0712.1767.

\bibitem{Giessbl} F. Giessibl, Rev. Mod. Phys. \textbf{75}, 949 (2003).

\bibitem{Boyer} L. Boyer {\it et al.}, J. Phys. D: Appl. Phys. \textbf{27}, 1504 (1994).

\bibitem{Kimthesis} W. J. Kim, {\sl Towards the experimental verification
  of macroscopic phenomena in quantum electrodynamics}, Ph. D. Thesis,
  Dartmouth College, August 2007; W.J. Kim {\it et al.}, in preparation.

\bibitem{OnoCaru} R. Onofrio and G. Carugno, Phys. Lett. A \textbf{198}, 365 (1995).

\bibitem{Cappella} B. Cappella and G. Dietler, Surf. Sci. Rep. \textbf{34}, 1 (1999).

\bibitem{Garcia} R. Garc\'ia and R. P\'erez, Surf. Sci. Rep. \textbf{47}, 197 (2002).

\end{references}
\end{document}